\documentclass[twocolumn]{aastex631}
\shorttitle{Impacts of BH-forming SNe on the diffuse $\nu$ background}
\shortauthors{Nakazato et al.}
\begin{document}

\title{Impacts of Black-Hole-Forming Supernova Explosions on the Diffuse Neutrino Background}

\author[0000-0001-6330-1685]{Ken'ichiro Nakazato}
\affiliation{Faculty of Arts and Science, Kyushu University, Fukuoka 819-0395, Japan}

\author[0000-0002-9234-813X]{Ryuichiro Akaho}
\affiliation{Faculty of Science and Engineering, Waseda University, Tokyo 169-8555, Japan}

\author[0000-0003-4136-2086]{Yosuke Ashida}
\affiliation{Department of Physics and Astronomy, University of Utah, Salt Lake City, UT 84112, USA}

\author[0000-0002-9397-3658]{Takuji Tsujimoto}
\affiliation{National Astronomical Observatory of Japan, Mitaka, Tokyo 181-8588, Japan}



\begin{abstract}

Flux spectrum, event rate, and experimental sensitivity are investigated for the diffuse supernova neutrino background (DSNB), which originates from past stellar collapses and is also known as a supernova relic neutrino background. For this purpose, the contribution of collapses that lead to successful supernova (SN) explosion and black hole (BH) formation simultaneously, which are suggested to be a non-negligible population from the perspective of Galactic chemical evolution, is taken into account. If the BH-forming SNe involve the matter fallback onto the protoneutron star for the long term, their total emitted neutrino energy becomes much larger than that of ordinary SNe and failed SNe (BH formation without explosion). Then, in the case of the normal mass hierarchy in neutrino oscillations and with half of all core-collapse SNe being BH-forming SNe, the expected event rate according to the current DSNB model is enhanced by up to a factor of two due to the BH-forming SNe. While substantial uncertainties exist regarding the duration of the matter fallback, which determines the total amount of emitted neutrinos, and the fraction of BH-forming SNe, the operation time required to detect the DSNB at Hyper-Kamiokande would be reduced by such contribution in any case.
%\footnote{Abstracts for the Astrophysical Journal (ApJ), the Astrophysical Journal Letters (ApJL), the Astronomical Journal (AJ), and the Planetary Science Journal (PSJ) all have a 250 word limit}.
\end{abstract}


\section{Introduction} \label{sec:intro}
Supernova (SN) explosions supply the elements synthesized inside stars, serving as the building blocks for the next generation of stars, planets, and life. Understanding the properties of progenitors and explosion mechanisms of SNe is crucial for unraveling the history of the Universe. In particular for the study of core-collapse SNe, neutrinos are expected to be a powerful tool. Stars with masses larger than $\sim$8$M_\odot$ experience core collapse at the end of their evolution, leading to the emission of a huge amount of neutrinos. In fact, neutrinos from SN1987A, which is a core-collapse SN that appeared in the Large Magellanic Cloud, were successfully detected \citep[][]{1987PhRvL..58.1490H,1987PhRvL..58.1494B,1987JETPL..45..589A}. In case that a next Galactic SN occurs, currently operating neutrino detectors, such as Super-Kamiokande (SK), are expected to detect a statistically sufficient number of neutrino events \citep[e.g.,][]{2011A&A...535A.109A,2016APh....81...39A,2024arXiv240306760K} and various insights into SN explosions will be derived from the neutrino observations \citep[][]{2021ApJ...916...15A,2022ApJ...934...15S,2024arXiv240418248S,2022MNRAS.512.2806N,2023ApJ...954...52H}.

Another approach to detecting neutrinos from SNe is to concentrate on the cosmic background. Neutrinos from distant SNe have been accumulated to form the diffuse SN neutrino background (DSNB), or SN relic neutrino (SRN) background. The upper bounds on the $\bar\nu_e$ flux have been provided by detectors such as SK \citep[][]{2012PhRvD..85e2007B,2015APh....60...41Z,2021PhRvD.104l2002A}, KamLAND \citep[][]{2012ApJ...745..193G,2022ApJ...925...14A}, and SK-Gd \citep[][]{2023ApJ...951L..27H}, which is the SK experiment with gadolinium-loaded water \citep[][]{2004PhRvL..93q1101B}. Furthermore, a larger-volume water Cherenkov detector, Hyper-Kamiokande (HK), is currently under construction \citep[][]{2018arXiv180504163H}, and other types of detectors such as liquid scintillators and argon/xenon-based detectors are planned. Predictions of the DSNB detection are exhibited for these future detectors including JUNO and DUNE \citep[][]{2017JCAP...11..031P,2018JCAP...05..066M,2021PhRvD.103b3021S,2022Univ....8..181L,2022PhRvD.105d3008S}.

Theoretical estimations of the DSNB flux and predictions of the event rates have been investigated for a long time \citep[][]{1982SvA....26..132B,1984Natur.310..191K,1985PhRvL..55.1422D}. Since models of DSNB involve a wide range of physical and astrophysical factors, improvements to the model have been made in many aspects. The spectrum of neutrinos emitted from a core-collapse SN depends on the progenitor, particularly its initial mass and metallicity. For estimating the DSNB flux, the average spectrum weighted by the initial mass function (IMF) is often used \citep[][]{1995APh.....3..367T}, and a large sample of progenitor models is currently under consideration \citep[][]{2018MNRAS.475.1363H,2021ApJ...909..169K}. The variation of IMF is also being discussed \citep[][]{2022MNRAS.517.2471Z,2023ApJ...946...69A,2023ApJ...953..151A}. The evolution of progenitors may be affected by binary interactions \citep[][]{2021PhRvD.103d3003H}. The cosmic SN rate, or star formation history, is deduced from astronomical observations \citep[][]{1996ApJ...460..303T,1997APh.....7..137H,1997APh.....7..125M,2000PhRvD..62d3001K,2009PhRvD..79h3013H,2014ApJ...790..115M} and the cosmic chemical evolution contributes to the metallicity distribution of progenitors \citep[][]{2015ApJ...804...75N}. The emission of neutrinos from SNe itself involves uncertainty, particularly in the late phase \citep[][]{2013PhRvD..88h3012N,2022ApJ...937...30A,2022PhRvD.106d3026E,2024PhRvD.109b3024E}. Flavor mixing caused by neutrino oscillations is a factor that influences the event rates of DSNB \citep[][]{2003APh....18..307A,2010PhRvD..81e3002G}, and the effects of exotic physics in the neutrino sector are also being investigated \citep[][]{2003PhLB..570...11A,2004PhRvD..70a3001F,2020PhRvD.102l3012D,2022PhRvD.106j3026D,2021JCAP...05..011T,2023PhRvD.107b3017I}. Furthermore, the DSNB flux may be related to other cosmic background radiation such as MeV $\gamma$-rays \citep[][]{2005JCAP...04..017S,2023ApJ...950...29A} and non-thermal high-energy neutrinos \citep[][]{2024arXiv240112403A}. Incidentally, neutrinos emitted from accretion disks formed around SNe may contribute to the cosmic background radiation \citep[][]{2019PhRvD.100d3008S,2024ApJ...966..101W}. The basics of DSNB are covered in several previous reviews \citep[][]{2004NJPh....6..170A,2010ARNPS..60..439B,2016APh....79...49L,2020MPLA...3530011M,2023PJAB...99..460A}.

In the present paper, we focus on the stellar core collapses which lead to black hole (BH) formation. Neutrinos are emitted from the BH-forming collapse, as well as the ordinary core-collapse SNe, and these neutrinos are in the same energy regime as the DSNB \citep[][]{2005APh....23..303I,2006ApJ...645..519N,2009PhRvL.102w1101L}. In previous studies on the DSNB, two scenarios for the core collapse of massive stars are considered: those resulting in an ordinary SN explosion, leaving a neutron star (NS), and those leading to BH formation without an explosion, which are known as failed SNe. This limitation on the scenario of core collapse makes sense because, according to numerical studies, successful explosions that make BHs are predicted as a relatively rare event \citep[see Fig.~13 of][]{2016ApJ...821...38S}. Recently, however, the examples for this case have been investigated in detail \citep[e.g.,][]{2023ApJ...957...68B}. Furthermore, some astronomical phenomena such as GRB 980425/SN1998bw \citep{1998Natur.395..672I} and W50/SS 433 \citep{2007MNRAS.377.1187P} are likely to be associated with core-collapse SNe that result in the formation of BHs. Hereafter, we refer to this subset as BH-forming SNe.

The perspective of nucleosynthesis implied from chemical evolution in the early Galaxy has significantly raised the importance of contribution from BH-forming SNe. This suggests two types of BH-forming SNe, each of which is implied to be a non-negligible population \citep[e.g.,][]{2006NuPhA.777..424N}. First, some studies propose that core-collapse SNe with large explosion energy ($\gtrsim$10$^{52}$~erg), which are referred to as hypernovae, should promote chemical enrichment with its contributed fraction as much as 50\% of massive ($>$ 20 $M_\odot$) stars to account for the observed abundance of some heavy elements including Zn among Galactic halo stars \citep[][]{2006ApJ...653.1145K,2020ApJ...900..179K}. Based on their model, the fraction of hypernovae corresponds to approximately 17\% of SNe and a much lower fraction would fail to explain the observed chemical evolution of several elements, such as Zn. Conversely, higher estimates of as much as 30\% are acceptable from the perspective of nucleosynthesis and chemical evolution, though such a high fraction may seem less realistic based on current understanding. Therefore, about 30\% would be regarded as an upper bound for the fraction of hypernovae. These hypernovae are generally considered to be BH-forming SNe that generate a jet-like explosion powered by energy from the rotating BH. In addition, hypernovae could play a key role as the dominant site of $\nu p$-process nucleosynthesis that can account for a large portion of abundances of Mo and Ru for low-metallicity stars \citep[][]{2015ApJ...810..115F,2022ApJ...924...29S}.

Another candidate of BH-forming SNe are the so-called faint SNe whose luminosities are very low due to a negligibly small ejection of synthesized $^{56}$Ni \citep[e.g.,][]{2003ApJ...591..288H}. Faint SNe have been highlighted as the origin of a subset of carbon-enhanced metal-poor (CEMP) stars in the Galactic halo \citep{2013ARA&A..51..457N}. An event frequency of faint SNe could be approximately inferred from the fraction of this kind of stars against halo stars, which is estimated to be 20\% of all SNe \citep{2014ApJ...797...21P}. Each of the low-metallicity stars, such as CEMP stars, is assumed to have been born from an individual SN. This is justified by the fact that each elemental abundance of stars in this metallicity range is compatible with that of each SN nucleosynthesis pattern. Nevertheless, it appears that at least some of the CEMP stars can be explained by the binary scenario (accretion of material enriched in C by AGB stars) rather than faint SNe \citep[e.g.,][]{2005ApJ...625..825L}. Therefore, 20\% is considered as an upper bound for the fraction of faint SNe. In addition to these arguments, there is a recent report that the inclusion of contribution from faint SNe with tens of percent leads to better agreement with the observed abundance of stars in the solar neighborhood \citep{2023MNRAS.524.6295P}.

In the end, BH-forming SNe that possibly emerge as hypernovae or faint SNe, are suggested to be a non-minor population that could be, as not an unrealistic case, counted up to a half of all core-collapse SNe, where 30\% from hypernovae and 20\% from faint SNe. However, it is noted that these estimates include large uncertainties and 50\% represents an optimistic upper bound. Further observational and theoretical investigations are important.

Although the detailed explosion mechanisms and formation processes of BHs are not identified for these BH-forming SNe, they should accompany the emission of a huge amount of neutrinos as well as the ordinary core-collapse SNe and failed SNe. In this study, we investigate the impact of BH-forming SNe on the DSNB flux updating our previous study \citep[][]{2023ApJ...953..151A}. This paper is organized as follows. In \S~\ref{sec:fbibh}, we describe the spectral model of neutrinos emitted from the BH-forming SNe. While the dynamics of BH-forming SNe includes uncertainties, we focus on the case induced by fallback mass accretion with a long duration. The formulation of the DSNB flux is presented in \S~\ref{sec:flux}. Issues concerning rates of core-collapse SNe and failed SNe based on the model of Galactic chemical evolution are also provided. In \S~\ref{sec:experim}, we investigate the DSNB event rate of $\bar\nu_e$. Furthermore, the experimental sensitivities at HK are evaluated. Finally, a summary and discussion are provided in \S~\ref{sec:discuss}.

\section{Neutrino emission from fallback induced BH formation} \label{sec:fbibh}
In this section, we consider the neutrino emission from the BH-forming SNe. Unfortunately, their dynamics is still not well understood. In particular, the time interval from the core bounce to BH formation, which corresponds to the duration of the neutrino emission, has notable impacts; generally the total energy of emitted neutrinos gets larger for a longer duration \citep{2021ApJ...909..169K}. To address this, we assume two extreme cases: one where a BH is formed dynamically on short timescales of $O(1)$~s, and another where the moderate fallback causes the mass of the protoneutron star (PNS) to exceed the maximum mass, resulting in the formation of a BH at later stages on timescales of $> \! O(10)$~s. Since the former case is similar to that of the failed SNe, we adopt the same spectral model of the emitted neutrinos as the failed SNe for the prompt BH-formation case. On the other hand, as for the case of fallback induced BH formation, we construct a neutrino spectrum in the present paper. In the following, we describe the model of emitted neutrinos from fallback induced BH formation. 

We combine the model of stellar core collapse in \citet{2021PASJ...73..639N} and the model of fallback mass accretion in \citet{2024ApJ...960..116A} for the evaluation of the neutrino spectrum from the fallback induced BH formation. In \citet{2024ApJ...960..116A}, the neutrino luminosity emitted by fallback mass accretion onto a PNS with the gravitational mass of $1.98M_\odot$, whose baryon mass corresponds to $2.35M_\odot$, is provided.

So as to estimate the neutrino spectrum emitted during the early dynamical phase, we utilize the core-collapse simulation of a $30M_\odot$ progenitor in \citet{2021PASJ...73..639N}. We integrate the neutrino emission of this model up to the point where a baryon mass of the PNS reaches $2.35M_\odot$. While three models with different nuclear equations of state (EOS) are presented in \citet{2021PASJ...73..639N}, we adopt the model with the Togashi EOS \citep{2017NuPhA.961...78T} in this study. Note that, the Furusawa--Togashi EOS \citep{2017JPhG...44i4001F} is used in the fallback model of \citet{2024ApJ...960..116A} and it is different from the Togashi EOS in the low-density regime. Nevertheless, the impact of their difference is minor on the neutrino emission during the early dynamical phase \citep{2023PTEP.2023a3E02S}.

In the fallback mass accretion model of \citet{2024ApJ...960..116A}, the inside PNS is assumed to have an isotropic temperature of $T = 2$~MeV. In contrast, the temperature of the PNS is $T \sim O(10)$~MeV when the baryon mass reaches $2.35M_\odot$ in \citet{2021PASJ...73..639N}. Therefore, in order to bridge the temperature gap, we perform the cooling simulation of the PNS and assess the neutrino emission. For this purpose, we adopt the structure of PNS obtained by the core-collapse simulation in \citet{2021PASJ...73..639N} as the initial condition. As for the EOS, Furusawa--Togashi EOS is adopted. Since the PNS in the initial condition is composed of uniform nuclear matter, for which the Furusawa--Togashi EOS uses the same model as the Togashi EOS, the impact of their differences is minor \citep{2023PTEP.2023a3E02S}. Our numerical methods for the PNS cooling are similar to those employed in \citet{1994pan..conf..763S} and \citet{2013ApJS..205....2N}. As a result of the simulation, we find that the PNS cools to $T \sim 2$~MeV over a period of 200~s. Since the effect of convection is not taken into account in this evaluation, the cooling time might be shorter. Nevertheless, since the total energy of neutrinos emitted from the PNS cooling stems from the binding energy of PNS, the time-integrated spectrum is insensitive to the cooling time.

We also incorporate the neutrinos emitted from the fallback mass accretion based on \citet{2024ApJ...960..116A}. Here, we consider the fallback mass accretion of $0.002M_\odot \, {\rm s}^{-1}$ onto a PNS with the gravitational mass of $1.98M_\odot$. Since, based on their adopted EOS, the maximum baryon mass of the NSs is $2.70M_\odot$ and the baryon mass of our PNS is $2.35M_\odot$, it takes 175~s to form a BH by the accretion of $0.35M_\odot$. Thus, the total amount of neutrinos emitted from the fallback mass accretion would be evaluated by integrating the emission rate of this model over 175~s. However, it is noted that, since the neutrino luminosity shown in \citet{2024ApJ...960..116A} includes the emission from the PNS in a quasi-steady state, we should subtract its contribution to avoid duplication from the contribution of PNS cooling evaluated above. As shown in Fig.~7 of \citet{2024ApJ...960..116A}, the neutrino luminosity and the mass accretion rate exhibit a linear relationship with an offset. This offset can be evaluated by extrapolation and regarded as the emission from the PNS. Therefore, we subtract the offset spectrum to incorporate the net contribution of the fallback mass accretion. Owing to the subtraction, the total amount of neutrinos emitted from the fallback mass accretion  is insensitive to the choice of mass accretion rate, provided that the accreted mass is fixed to $0.35M_\odot$.

\begin{figure*}[tbp]
  \gridline{\fig{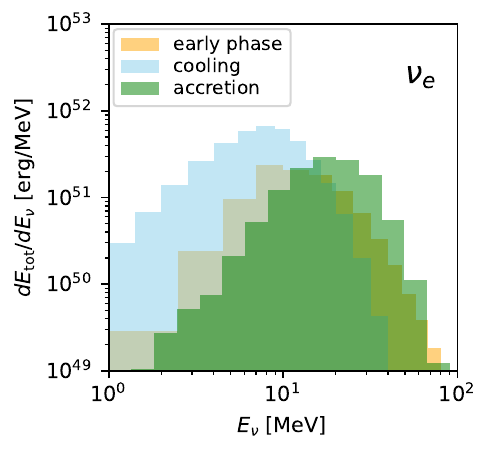}{0.33\textwidth}{(a) $\nu_e$}
            \fig{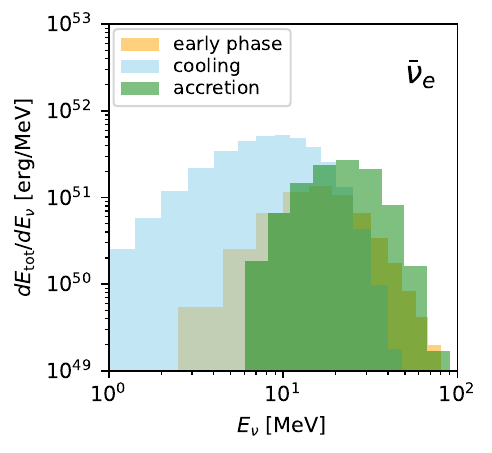}{0.33\textwidth}{(b) $\bar\nu_e$}
            \fig{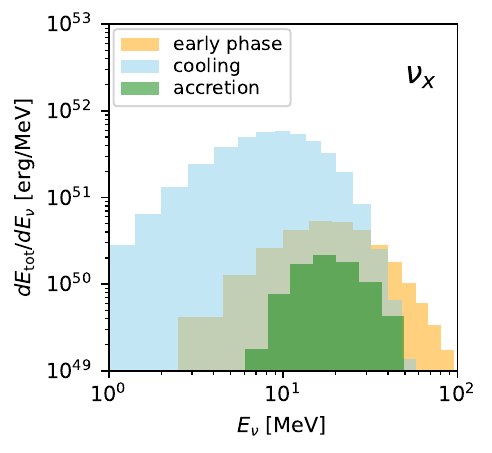}{0.33\textwidth}{(c) $\nu_x$}}
  \vspace{+5truept}
  \caption{Time integrated spectra of (a) $\nu_e$, (b) $\bar\nu_e$, and (c) $\nu_x$, where $\nu_x=\nu_\mu=\bar\nu_\mu=\nu_\tau=\bar\nu_\tau$, for the fallback induced BH formation model. Orange, blue, and green histograms correspond to the components of the early dynamical phase, the PNS cooling, and the fallback mass accretion, respectively.}
  \label{fig:spectcomp}
  \end{figure*}

In Fig.~\ref{fig:spectcomp}, we show the neutrino spectra of individual components: the early dynamical phase, the PNS cooling, and the fallback mass accretion. A substantial amount of $\nu_e$ and $\bar\nu_e$ are emitted from the fallback mass accretion. These neutrinos have high average energies and they are mainly produced by electron capture and positron capture. Fallback mass accretion produces a high-temperature environment where thermal electrons and positrons are created and supplies enormous free protons and neutrons which capture electrons and positrons, respectively. In contrast, the amount of $\nu_x$ ($=\nu_\mu=\bar\nu_\mu=\nu_\tau=\bar\nu_\tau$) emitted from the fallback mass accretion is much smaller than those of $\nu_e$ and $\bar\nu_e$. According to \citet{2024ApJ...960..116A}, the primary process for emitting $\nu_x$ is nucleon--nucleon bremsstrahlung, which is efficient in the high-density regions inside the PNS, and the luminosity of $\nu_x$ is hardly dependent on the accretion rate. Therefore, due to the subtraction considered in this study, the contribution of fallback mass accretion is minor for $\nu_x$.

In Table~\ref{tab:snnmodel}, the average and total energies of neutrinos emitted from the fallback induced BH-forming SN are compared with other scenarios taken into account in this study. The total emission energy of neutrinos, summing all flavors, amounts to $8.63\times 10^{53}$~erg. Since the binding energy of the maximum-mass NS is $8.78\times 10^{53}$~erg for the EOS adopted here, our fallback induced BH formation can be interpreted as maximizing the emission energy of neutrinos from a single stellar collapse\footnote{Except for the collapse of supermassive stars with $\gtrsim$10$^3M_\odot$ \citep[e.g.,][]{2006ApJ...645..519N}.}. Furthermore, while we combine the contributions of the early dynamical phase, the PNS cooling, and the fallback mass accretion in this study, this approach can be evaluated favorably from the perspective of energy conservation. On the other hand, as already mentioned, we assume that the neutrino signal of the prompt BH-forming SN is the same as the failed SN (BH formation without explosion). The total emission energy of neutrinos, summing all flavors, is $2.37\times 10^{53}$~erg, which is 28\% of the fallback induced BH formation model. Then, we consider that the uncertainty in the neutrino emission from BH-forming SNe is accounted for by both extremes.

%%%
%%%
%%%
  \begin{deluxetable*}{lccrrrrrr}
  \tablecaption{Properties of emitted neutrinos for the models considered in this study\label{tab:snnmodel}.}
  \tablewidth{0pt}
  \tablehead{
  \colhead{} & \colhead{} & \colhead{} & \colhead{$\langle E_{\nu_e} \rangle$} & \colhead{$\langle E_{\bar \nu_e} \rangle$} & \colhead{$\langle E_{\nu_x} \rangle$} & \colhead{$E_{\nu_e, {\rm tot}}$} & \colhead{$E_{\bar \nu_e, {\rm tot}}$} & \colhead{$E_{\nu_x, {\rm tot}}$} \\
  \colhead{Model} & \colhead{Explosion} & \colhead{Remnant} & \multicolumn3c{(MeV)} & \multicolumn3c{($10^{52}$ erg)}
  }
  \startdata
  Ordinary core-collapse SN & successful & NS & 9.2 & 10.9 & 11.8 & 4.47 & 4.07 & 4.37 \\
  BH-forming SN, (i) fallback induced & successful & BH &11.8 & 13.6 & 10.9 & 19.48 & 18.50 & 12.07 \\
  BH-forming SN, (ii) prompt & successful & BH & 16.1 & 20.4 & 23.4 & 6.85 & 5.33 & 2.89 \\
  Failed SN & failed & BH & 16.1 & 20.4 & 23.4 & 6.85 & 5.33 & 2.89 \\
  \enddata
  \tablecomments{$\langle E_{\nu_i} \rangle$ and $E_{\nu_i, {\rm tot}}$ are the average and total energies of the time-integrated neutrino signal for $\nu_i$, where $\nu_x$ represents the average of $\nu_\mu$, $\bar \nu_\mu$, $\nu_\tau$, and $\bar \nu_\tau$.}
  \end{deluxetable*}
%%%

\section{DSNB Flux Model} \label{sec:flux}
Following \citet{2023ApJ...953..151A}, we describe the DSNB flux as
  \begin{eqnarray}
    & & \frac{d\Phi(E_\nu)}{dE_\nu}=  c\int^{z_{\rm max}}_{0}  \frac{dz}{H_0\sqrt{\Omega_{\rm m}(1+z)^3+\Omega_\Lambda}} \times \nonumber \\ & & \left[ R_{\rm SN}(z) \left\{ (1-f_{\rm BHSN})\frac{dN_{\rm CCSN}(E^\prime_\nu)}{dE^\prime_\nu} + \right. \right. \nonumber \\ & & \left. \left. f_{\rm BHSN}\frac{dN_{\rm BHSN}(E^\prime_\nu)}{dE^\prime_\nu} \right\} + R_{\rm BH}(z) \frac{dN_{\rm BH}(E^\prime_\nu)}{dE^\prime_\nu} \right],
  \label{eq:flux}
  \end{eqnarray}
where $c$ is the speed of light and the cosmological constants are $\Omega_{\rm m}= 0.3089$, $\Omega_\Lambda= 0.6911$, and $H_0= 67.74$~km~sec$^{-1}$~Mpc$^{-1}$. The neutrino energy at a detector, $E_\nu$, and a source, $E^\prime_\nu$, are related to the redshift of the source, $z$, as $E^\prime_\nu=(1+z)E_\nu$, where the range of redshift is set to $0\le z \le z_{\rm max} =5$. As for the neutrino sources, we consider ordinary core-collapse SNe, BH-forming SNe, and failed SNe, whose spectra are denoted as, $dN_{\rm CCSN}(E^\prime_\nu)/dE^\prime_\nu$, $dN_{\rm BHSN}(E^\prime_\nu)/dE^\prime_\nu$, and $dN_{\rm BH}(E^\prime_\nu)/dE^\prime_\nu$, respectively. For $dN_{\rm CCSN}(E^\prime_\nu)/dE^\prime_\nu$ and $dN_{\rm BH}(E^\prime_\nu)/dE^\prime_\nu$, we adopt the same models as \citet{2023ApJ...953..151A}, but we only investigate the case of the Togashi EOS. The model of an ordinary core-collapse SN corresponds to the case where a $1.32M_\odot$ NS is formed from the collapse of a $15M_\odot$ progenitor, while the model of a failed SN corresponds to the case where a BH is formed without an explosion from the collapse of a $30M_\odot$ progenitor \citep[][]{2022ApJ...937...30A}. As stated in \S~\ref{sec:fbibh}, we investigate two cases: (i) fallback induced and (ii) prompt BH-forming SNe (Table~\ref{tab:snnmodel}) for $dN_{\rm BHSN}(E^\prime_\nu)/dE^\prime_\nu$. 

In eq.~(\ref{eq:flux}), $R_{\rm SN}(z)$ and $R_{\rm BH}(z)$ are rates of successful SNe and failed SNe, respectively, as functions of the redshift. We adopt $R_{\rm SN}(z)$ and $R_{\rm BH}(z)$ deduced from the model of Galactic chemical evolution in \citet{2023MNRAS.518.3475T}, which are also investigated in \citet{2023ApJ...953..151A} as a reference model (see Table~1 of that paper). This model exhibits two distinct features. Firstly, the stellar IMF depends on the type of galaxies \citep[e.g.,][]{2018PASA...35...39H}; the early-type galaxies, which are formed in bursty star formation, have a flat IMF (a slope index of a power law $x=-0.9$) and efficiently eject heavy elements while late-type galaxies have the Salpeter IMF ($x=-1.35$). Secondly, the upper bound on the mass of core-collapse SN progenitors is $18M_\odot$ \citep[e.g.,][]{2009ARA&A..47...63S,2015PASA...32...16S,2016ApJ...821...38S,2021ApJ...909..169K}; a mass range of progenitors is 8--18$M_\odot$ for core-collapse SNe and 18--100$M_\odot$ for failed SNe. The predicted redshift evolution of $R_{\rm SN}$ is in better agreement with the measured rates. On the other hand, $R_{\rm BH}$ corresponds to the rate of BH formations without SN explosions. In the present study, core-collapse SNe are classified into two categories: ordinary core-collapse SNe, which leave NSs, and BH-forming SNe. For simplicity, we assume that the fraction of BH-forming SNe, denoted as $f_{\rm BHNS}$, does not depend on the redshift. In the following, so as to examine the impact of BH-forming SNe on the DSNB, we use $f_{\rm BHNS}=0.5$, which is an upper bound estimated in \S~\ref{sec:intro}, and $f_{\rm BHNS}=0.1$, which is a more moderate case, for illustration. However, in reality, $f_{\rm BHSN}$ is expected to be more complicated, involving the redshift dependence.

In \citet{2023MNRAS.518.3475T}, $R_{\rm SN}(z)$ and $R_{\rm BH}(z)$ are calculated by converting from the observationally estimated cosmic star formation rate (SFR) under the assumption that the progenitor-mass ranges of core-collapse SNe ($R_{\rm SN}$) and failed SNe ($R_{\rm BH}$) are fixed to 8--18$M_\odot$ and 18--100$M_\odot$, respectively. For this purpose, the SFRs of \citet{2014ARA&A..52..415M} and \citet{2006ApJ...651..142H}, which are referred to as MD14 and HB06, respectively, are used. We also investigate the both cases in this paper. 

Since neutrinos undergo flavor oscillations before the detection, we take the so-called MSW effect \citep[][]{1978PhRvD..17.2369W,1985YaFiz..42.1441M} into account following \citet{2015ApJ...804...75N}. The $\bar \nu_e$ survival probability $\bar P$ depends on the neutrino mass hierarchy \citep[][]{2000PhRvD..62c3007D} as $\bar P = \cos^2\theta_{12}\cos^2\theta_{13}$ for the normal mass hierarchy (NH) and $\bar P =\sin^2\theta_{13}$ for the inverted mass hierarchy (IH), where $\theta_{12}$ and $\theta_{13}$ are mixing angles. Since recent measurements for them are $\sin^2\theta_{12} \approx 0.31$ and $\sin^2\theta_{13} \approx 0.02$ \citep[][]{2022PTEP.2022h3C01W}, we set $\bar P = 0.68$ for NH and $\bar P = 0.02$ for IH in this study.

\begin{figure*}[tbp]
  \gridline{\fig{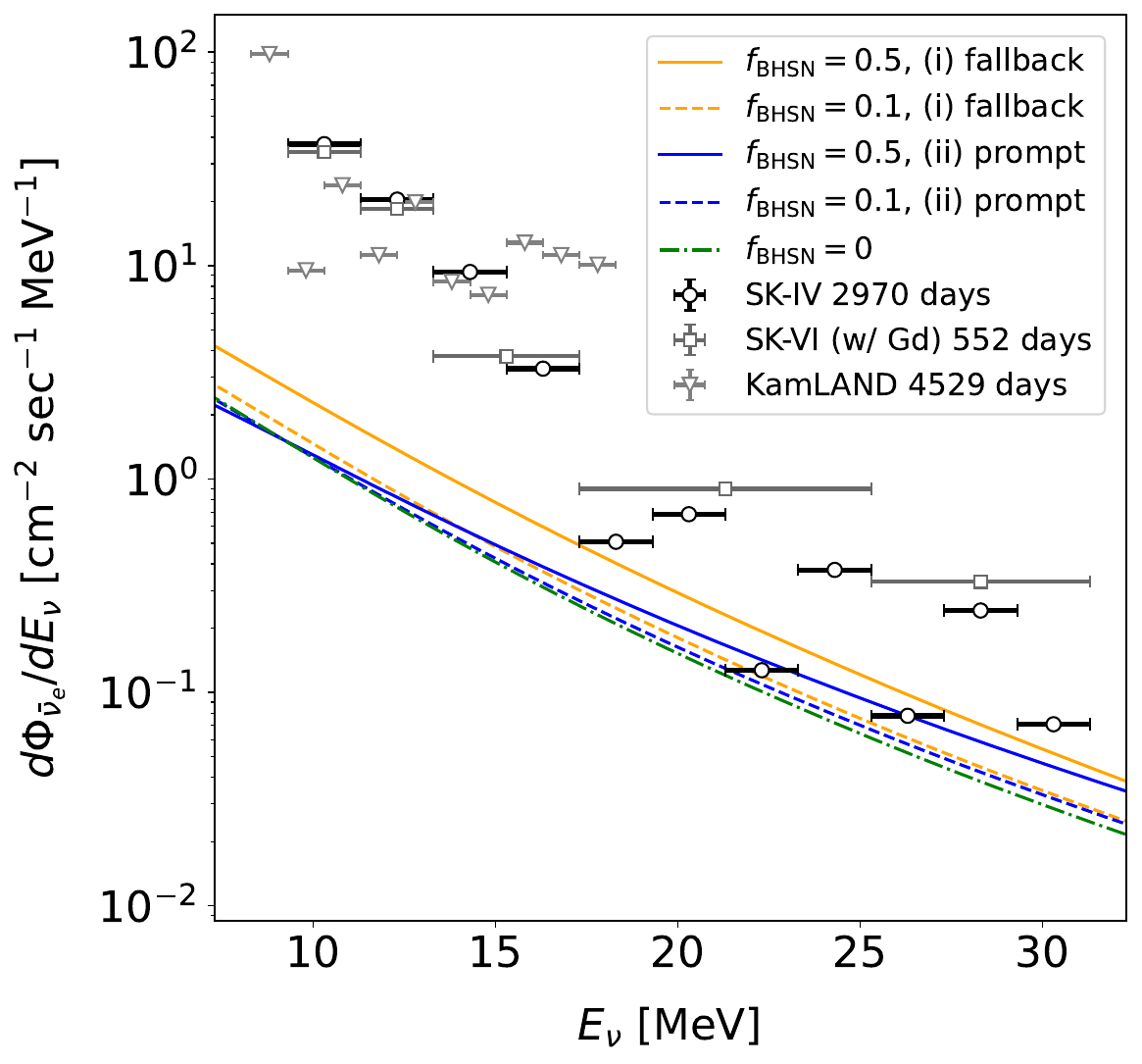}{0.48\textwidth}{(a) NH, MD14 SFR}\hspace{-15truept}
            \fig{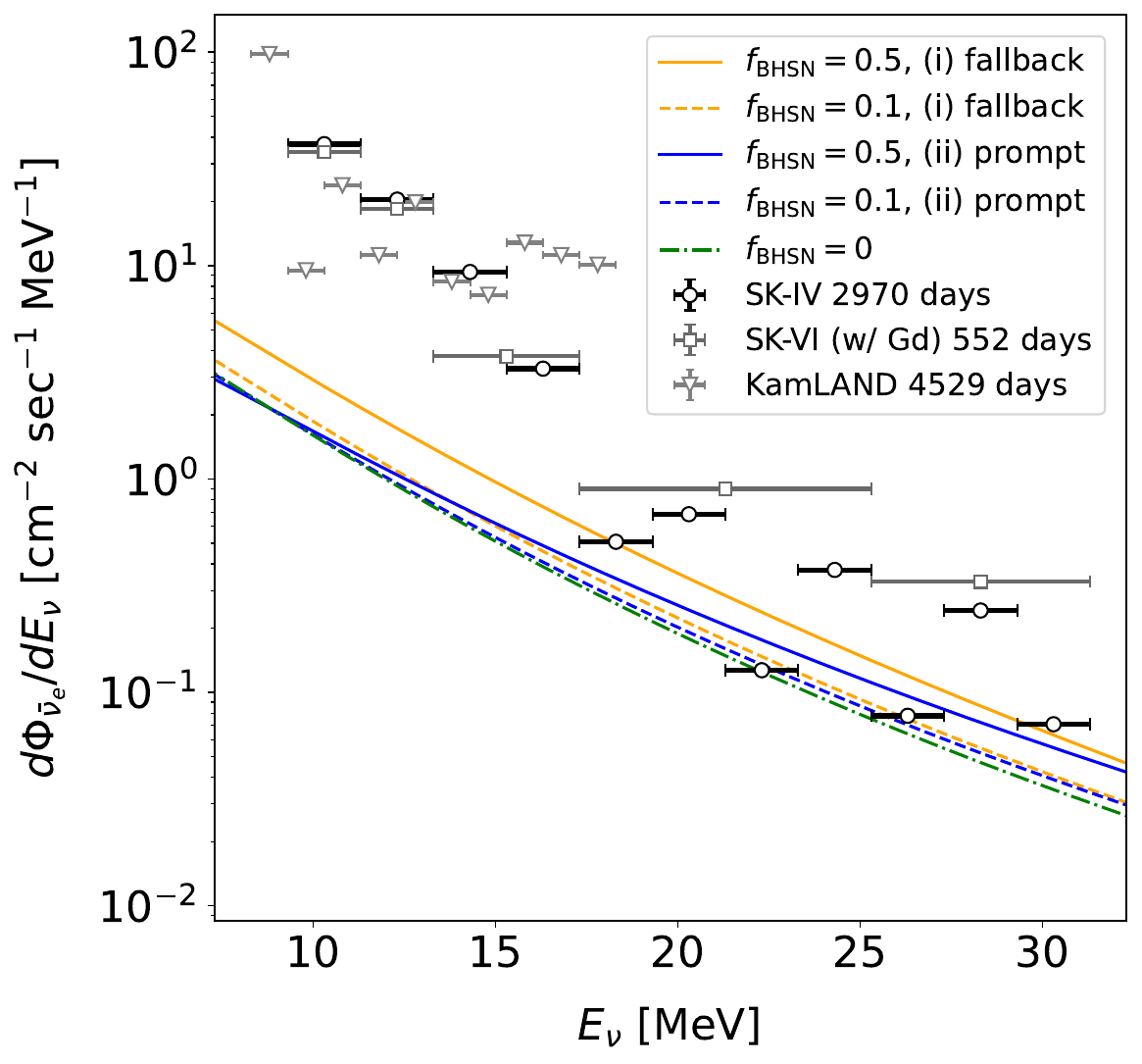}{0.48\textwidth}{(b) NH, HB06 SFR}}
  \gridline{\fig{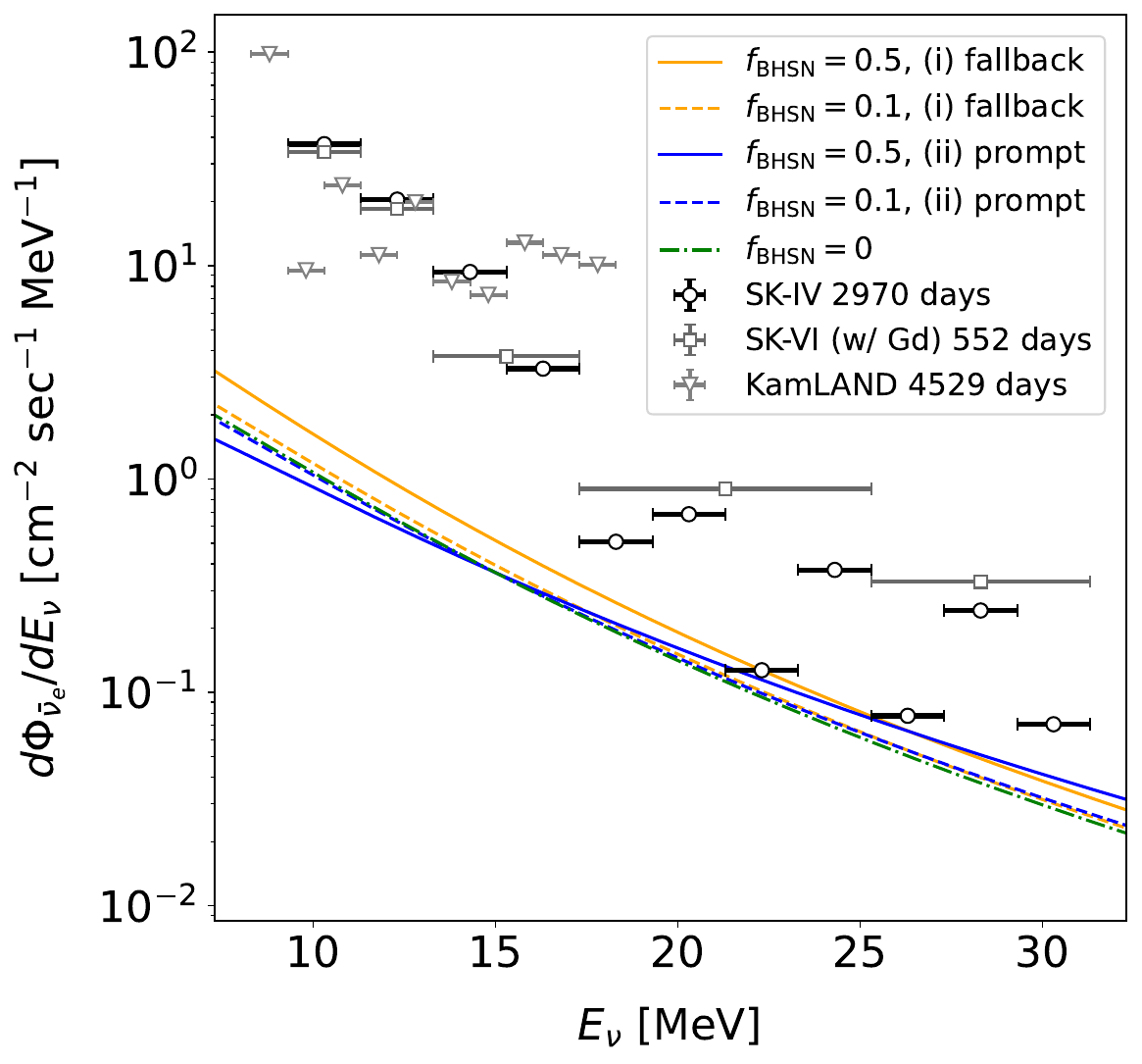}{0.48\textwidth}{(c) IH, MD14 SFR}\hspace{-15truept}
            \fig{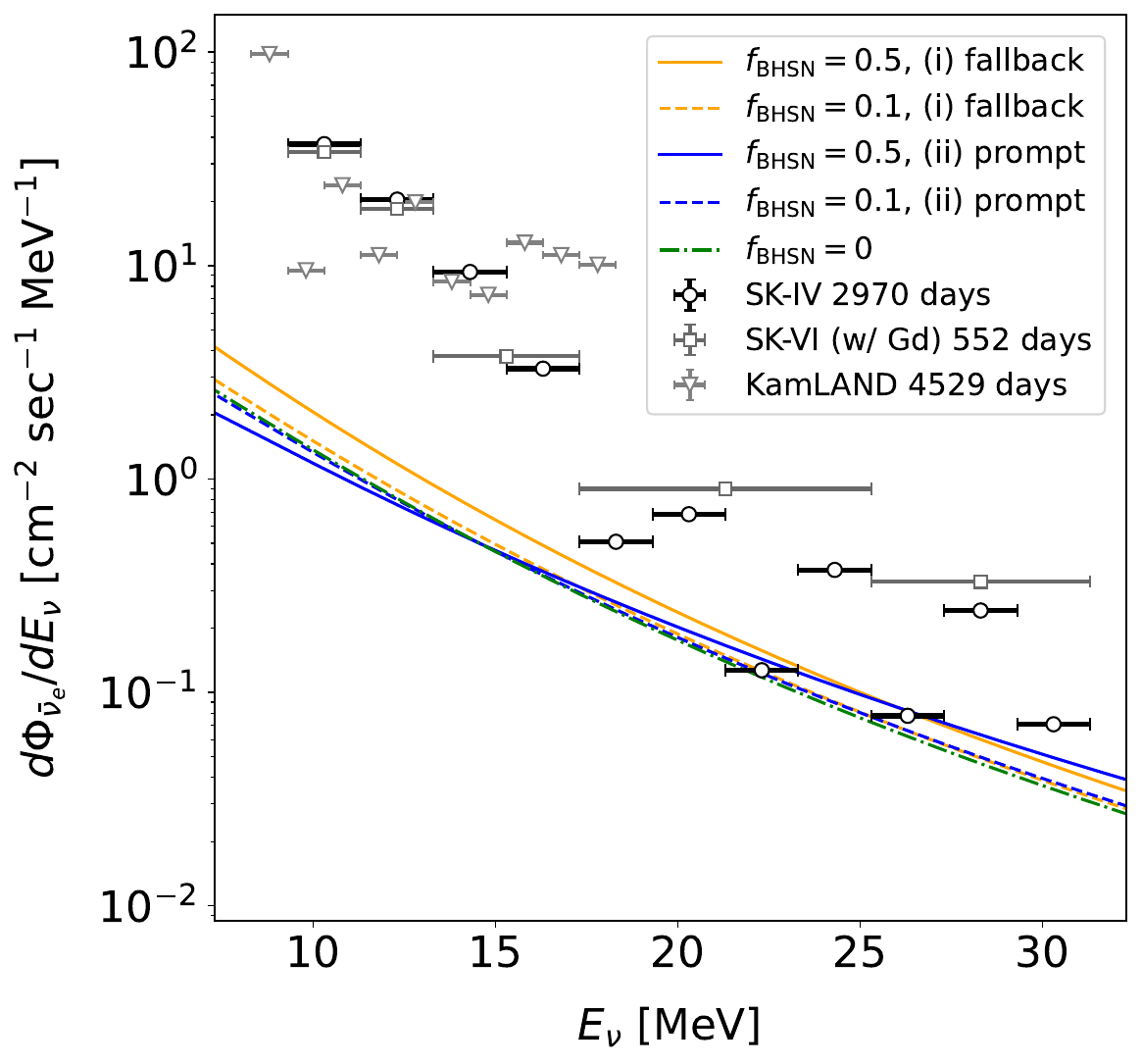}{0.48\textwidth}{(d) IH, HB06 SFR}}
  \vspace{+5truept}
  \caption{Spectra of cosmic background $\bar \nu_e$ flux from this study compared with the 90\% confidence level upper bounds from SK with pure \citep[][]{2021PhRvD.104l2002A} and gadolinium-loaded \citep[][]{2023ApJ...951L..27H} water, and KamLAND \citep[][]{2022ApJ...925...14A}. Models with different mass hierarchies and SFRs are shown in each panel: (a) NH and MD14, (b) NH and HB06, (C) IH and MD14, and (d) IH and HB06. Solid, dashed, and dot-dashed lines show the spectra estimated with $f_{\rm BHSN}=0.5$, 0.1, and 0, respectively, where $f_{\rm BHSN}$ is the fraction of successful SN explosions that form a BH. Orange and blue lines correspond to (i) fallback induced and (ii) prompt BH-formation cases, respectively, in Table~\ref{tab:snnmodel}.}
  \label{fig:srnfulx}
  \end{figure*}

\begin{figure}[tbp]
%  \gridline{\fig{fluxNHpub.pdf}{0.48\textwidth}{(a) NH}\hspace{-15truept}
%  \fig{fluxIHpub.pdf}{0.48\textwidth}{(b) IH}}
  \gridline{\fig{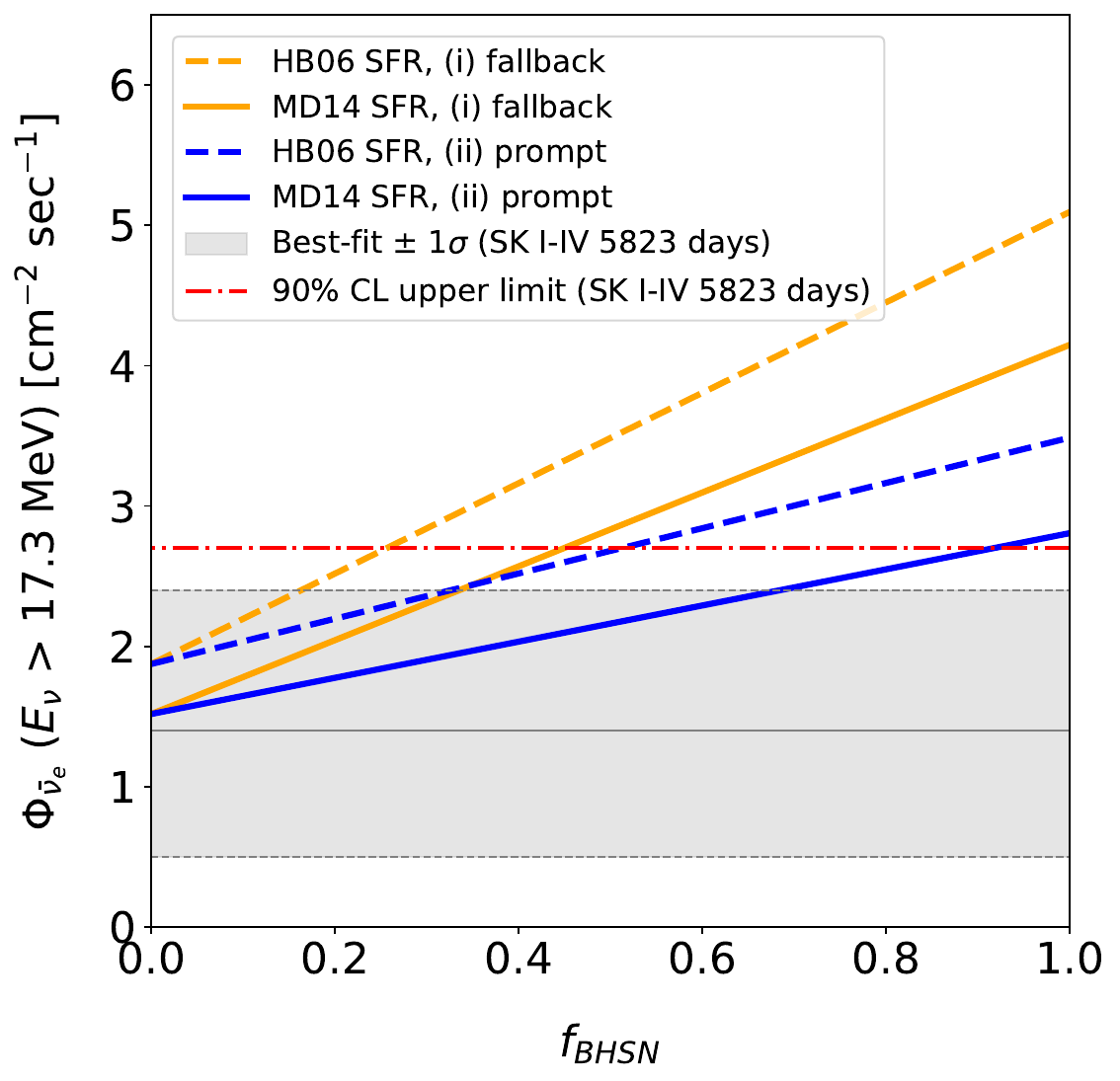}{0.45\textwidth}{(a) NH}}
  \gridline{\fig{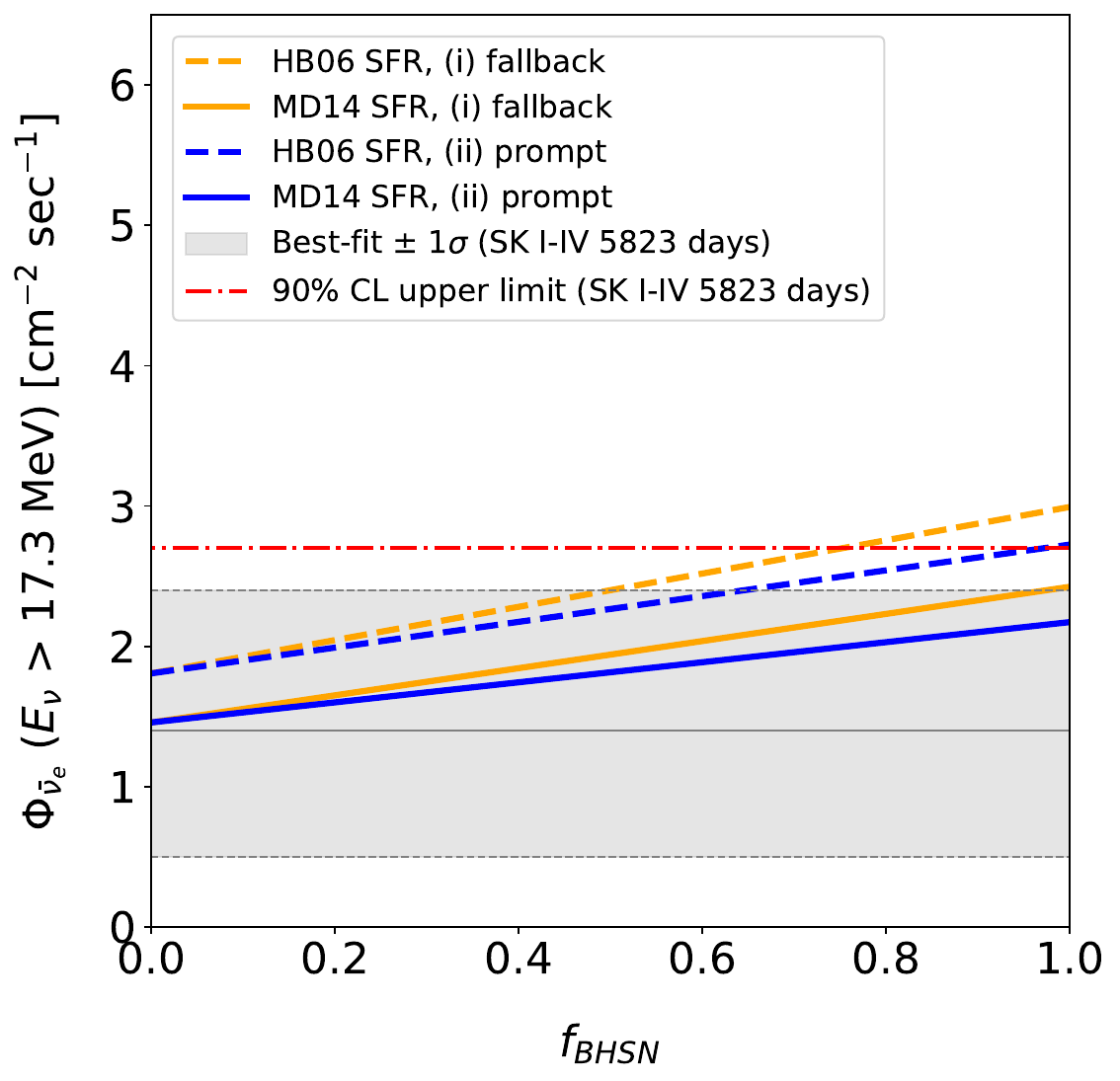}{0.45\textwidth}{(b) IH}}
  \vspace{+5truept}
  \caption{The integrated $\bar \nu_e$ flux with $E_\nu>17.3$~MeV estimated by our DSNB models in comparison with the best-fit values and their $\pm 1\sigma$ uncertainties shown in grey solid lines and shaded regions, respectively, and the 90\% observed upper limits shown in red dot-dashed lines from SK with pure water \citep[][]{2021PhRvD.104l2002A}. Note that these experimental constraints are associated with the spectral models in \citet{2015ApJ...804...75N} while they are insensitive to the adopted spectrum models. Models with different mass hierarchies are shown in each panel: (a) NH and  (b) IH. Solid and dashed lines represent the spectra for the models with MD14 SFR and HB06 SFR, respectively. Orange and blue lines correspond to (i) fallback induced and (ii) prompt BH-formation cases, respectively, in Table~\ref{tab:snnmodel}.}
  \label{fig:srnfulxcomp}
  \end{figure}

In Fig.~\ref{fig:srnfulx}, the $\bar\nu_e$ flux estimated by the DSNB model described above is compared with the latest experimental upper bounds. The largest flux is provided by the model with NH, HB06 SFR, and fallback induced BH formation for the following reason. In the case of fallback induced BH formation, the binding energy of the maximum-mass NS is converted to the total emission energy of neutrinos. Since, as already stated, the fallback mass accretion emits a much larger amount of $\bar\nu_e$ than $\nu_x$, the terrestrial DSNB flux is larger for NH, which has a higher survival probability of $\bar\nu_e$ than IH. If the fraction of BH-forming SNe is $f_{\rm BHSN}=0.5$ for NH, the DSNB flux exceeds the upper bounds in several energy bins. As shown in Fig.~\ref{fig:srnfulxcomp}, where the integrated fluxes with $E_\nu>17.3$~MeV are compared with the 90\% C.L. upper limits and best-fit results in \citet{2021PhRvD.104l2002A}, the models of fallback induced BH formation have a constraint of $f_{\rm BHSN}<0.45$ (0.26) for the case with NH and MD14 (HB06) SFR.

%%%
%%%
%
%%%
%%%

\section{Event Rate and Experimental Sensitivity} \label{sec:experim}
In this section, we investigate the event rate spectra and the experimental sensitivity to our DSNB models following \citet{2022ApJ...937...30A}. Water Cherenkov detectors, such as SK and HK, detect the DSNB via inverse beta decay (IBD) of $\bar\nu_e$:
  \begin{equation}
   \bar \nu_e + p \to e^+ + n .
  \label{eq:ibd}
  \end{equation}
Thus, the DSNB event rate is calculated as
  \begin{equation}
   \frac{dN_{\rm event}(E_{e^+})}{dE_{e^+}}=N_p\cdot\sigma_{\rm IBD}(E_\nu) \cdot\frac{d\Phi^{\rm det}_{\bar\nu_e}(E_\nu)}{dE_\nu},
  \label{eq:evrate}
  \end{equation}
where $\sigma_{\rm IBD}(E_\nu)$ is the IBD cross section taken from \citet{2003PhLB..564...42S} and $d\Phi^{\rm det}_{\bar\nu_e}(E_\nu)/dE_\nu$ is terrestrial flux of $\bar\nu_e$. The positron energy $E_{e^+}$ is related to the neutrino energy $E_\nu$ as $E_{e^+}=E_\nu-\Delta c^2$, where $\Delta$ is a neutron--proton mass difference, and $N_p$ represents the number of free protons contained in the fiducial volume of the detector, which is $N_p=1.5\times 10^{33}$ for SK and $N_p=12.6\times 10^{33}$ for HK.

\begin{figure}[tbp]
\centering
\includegraphics[width=0.475\textwidth]{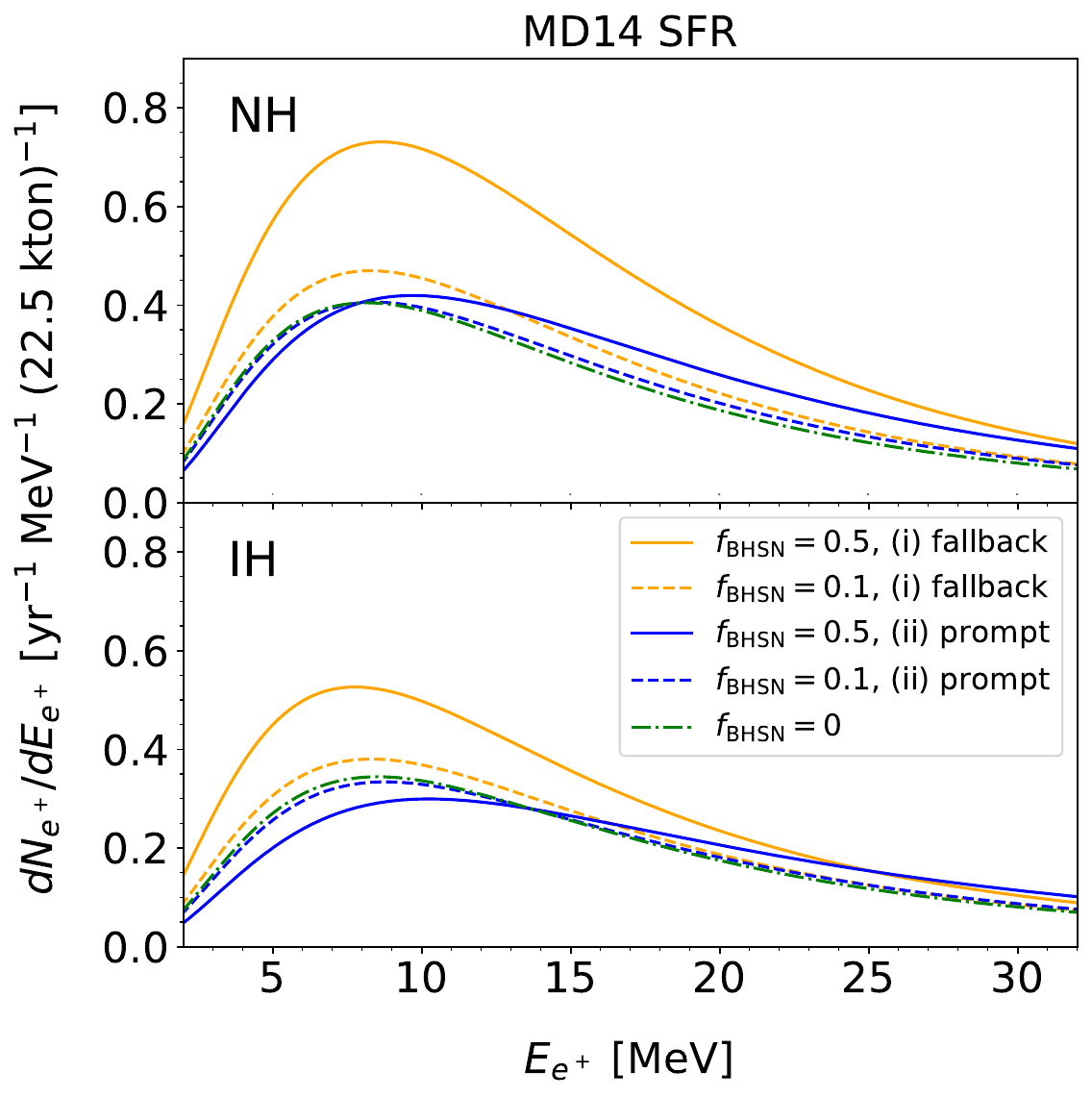}
\caption{Predicted DSNB event rate spectra at SK (a water volume of 22.5~kton) per year with different choices of the BH-forming SNe model and $f_{\rm BHSN}$ for NH (top) and IH (bottom) and MD14 SFR. The notation of lines is the same as in Fig.~\ref{fig:srnfulx}.}
\label{fig:evspect}
\end{figure}

The event rate spectra at SK from our DSNB models with MD14 SFR are shown in Fig.~\ref{fig:evspect}. Incidentally, HK has a $\sim$8.4 times higher event rate than SK and the model with HB06 SFR has a $\sim$1.24 times higher event rate than that with MD14 SFR. If there are no BH-forming SNe ($f_{\rm BHSN}=0$), the expected number of IBD signal events with $17.3<E_\nu<31.3$~MeV is 180 (170) for the model with NH (IH) and MD14 SFR at HK over 10 yr.\footnote{The event number is reduced to 50--60 when the detection efficiency including neutron tag is taken into account \citep[][]{2023ApJ...953..151A}.} If the contributions of fallback induced BH-forming SNe are included and $f_{\rm BHSN}=0.5$ is assumed, the event number increases to 340 for NH and 230 for IH. On the other hand, in the case of prompt BH-forming SNe and $f_{\rm BHSN}=0.5$, the event number is 260 for NH and 210 for IH. The impact on the event number is the largest for the case with NH and fallback induced BH-forming SNe. However, for the other cases also, the event numbers increase due to the BH-forming SNe while the impacts are not so large. This is because high-energy neutrino emission in the early dynamical phase is more efficient compared to ordinary core-collapse SNe. In any case, the inclusion of BH-forming SNe favors the detection of DSNB. Incidentally, it reduces the number of IBD signal events with $E_\nu<13.3$~MeV, where many background events exist at HK, for the case with IH and prompt BH-forming SNe. This is because low-energy neutrinos are mainly emitted from the cooling of the PNS, which is not included in the prompt BH formation case.

%%%
%%%

%%%
  \begin{figure*}[tbp]
  \gridline{\fig{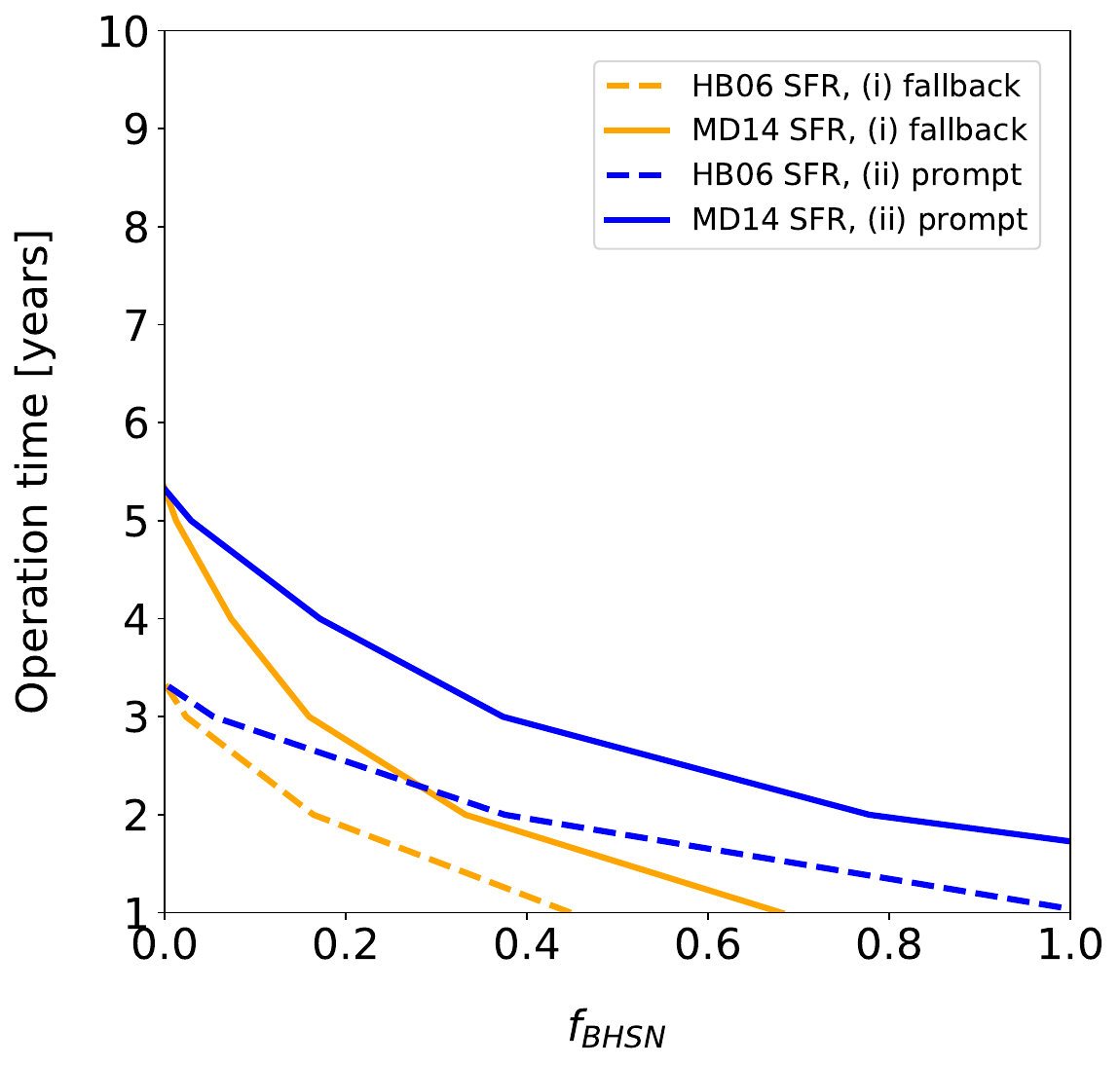}{0.48\textwidth}{(a) NH, 2$\sigma$}\hspace{-15truept}
            \fig{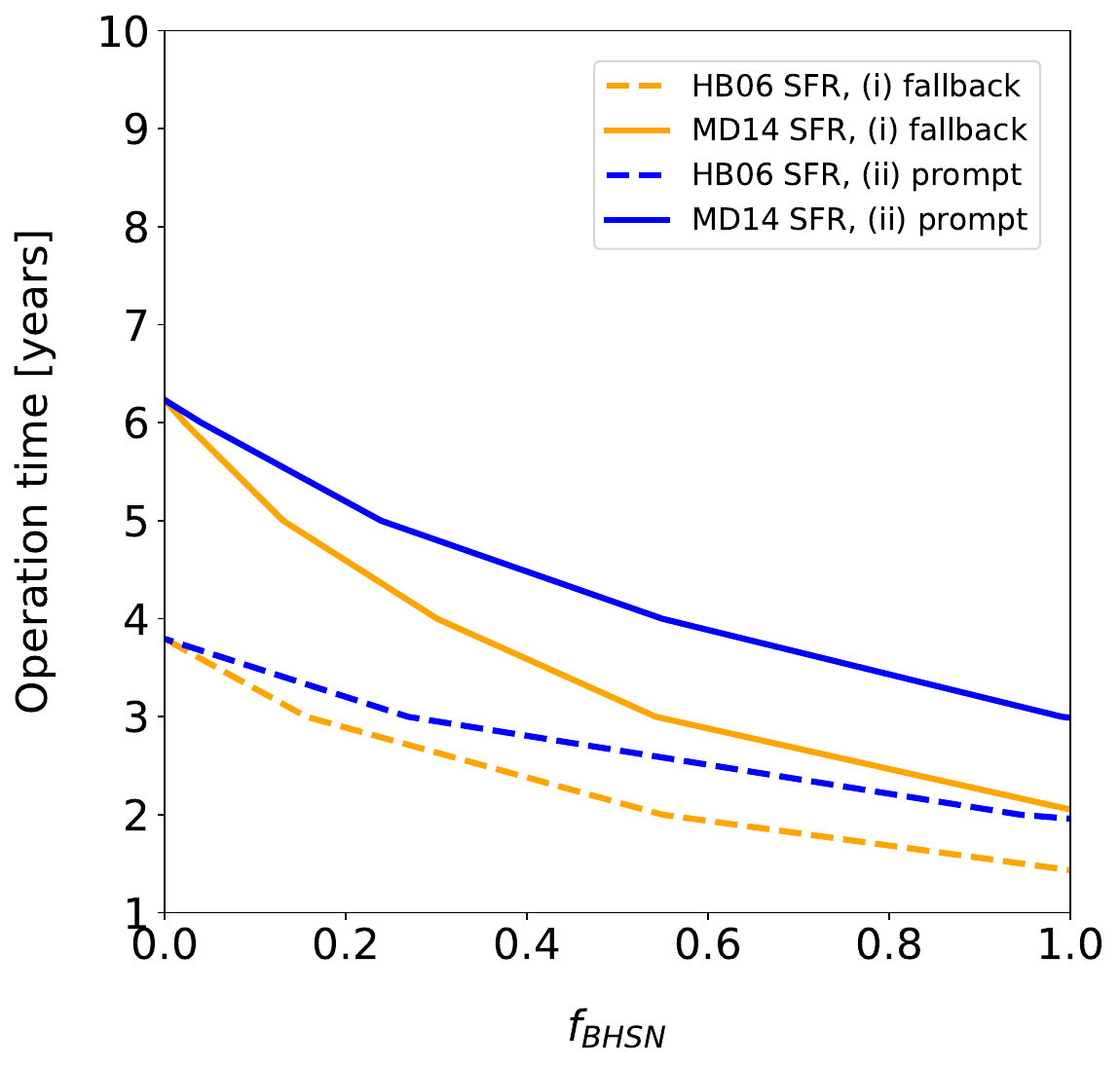}{0.48\textwidth}{(b) IH, 2$\sigma$}}
  \gridline{\fig{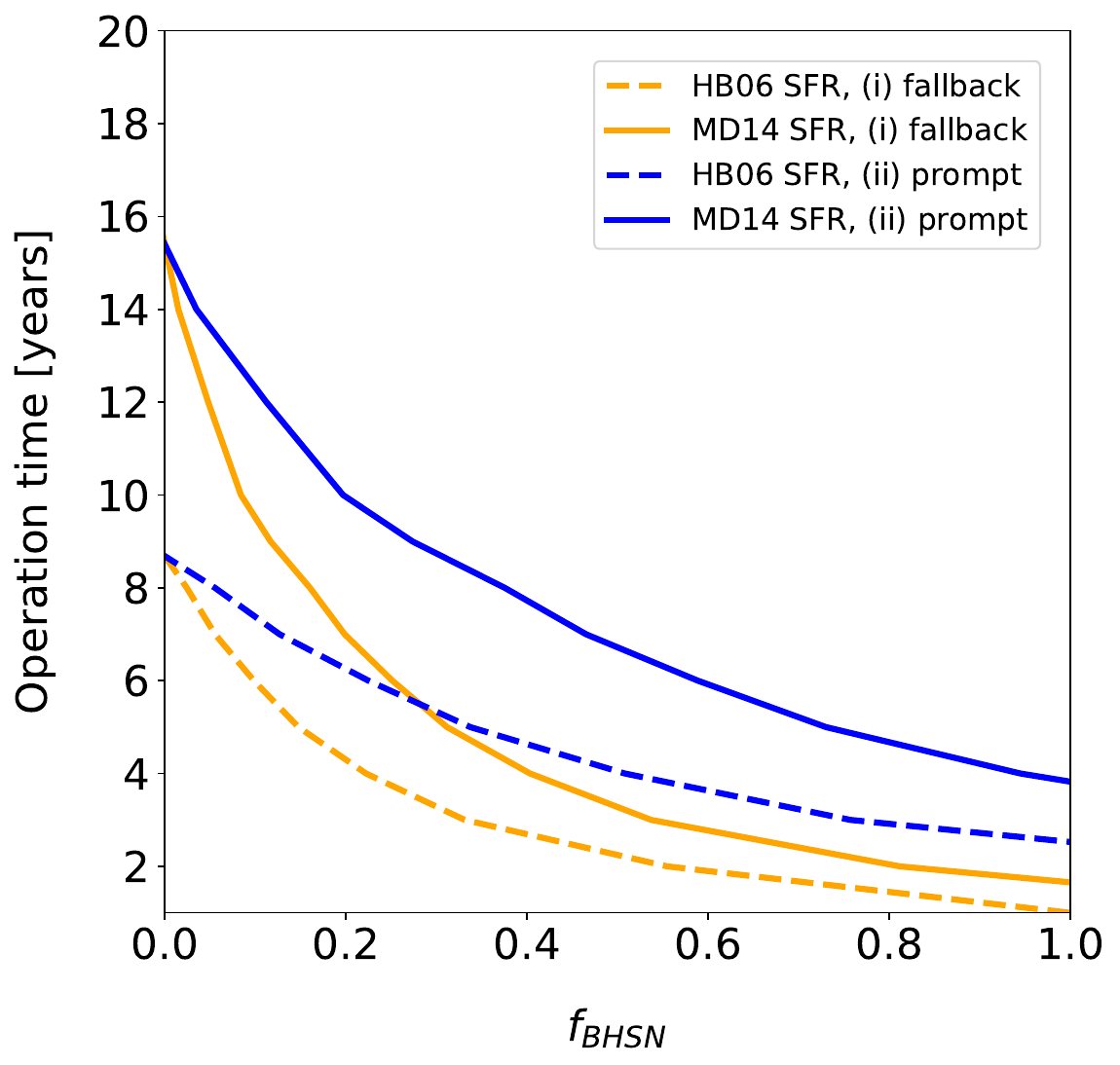}{0.48\textwidth}{(c) NH, 3$\sigma$}\hspace{-15truept}
            \fig{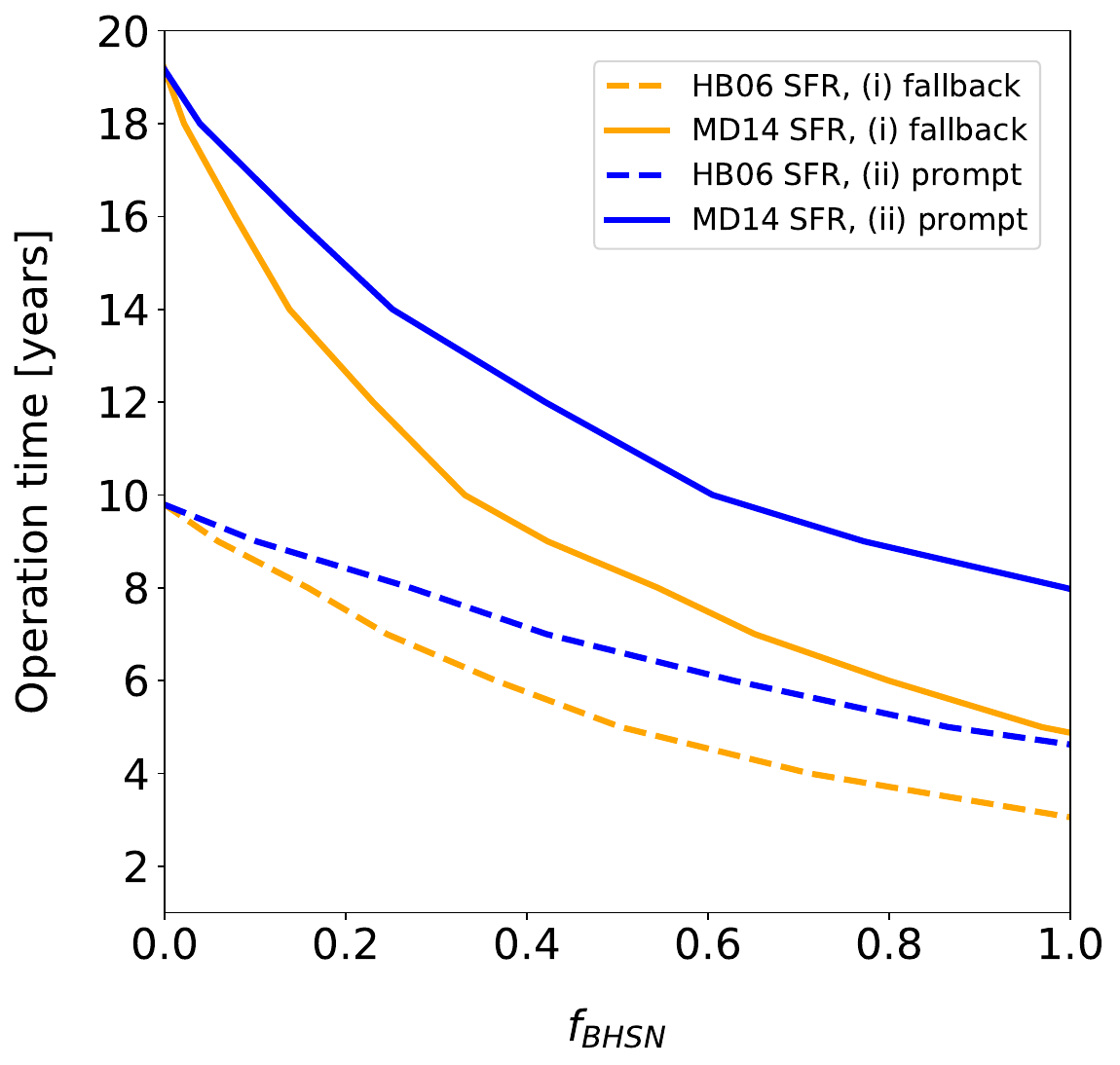}{0.48\textwidth}{(d) IH, 3$\sigma$}}
  \vspace{+5truept}
  \caption{Operation time required to detect the DSNB at HK as a function of the fraction of BH-forming SNe, $f_{\rm BHSN}$, based on our models with different mass hierarchy at different C.L.: (a) NH and 2$\sigma$, (b) IH and 2$\sigma$, (c) NH and 3$\sigma$, and (d) IH and 3$\sigma$. The notation of lines is the same as in Fig.~\ref{fig:srnfulxcomp}.}
  \label{fig:sensitivity}
  \end{figure*}
%%%
Now we move on to the experimental sensitivity. The expected upper bound on the integrated flux of $\bar\nu_e$ with $17.3<E_\nu<31.3$~MeV is calculated as \citep[][]{2021PhRvD.104l2002A,2022ApJ...937...30A}
  \begin{equation}
   \Phi_{\rm lim}=\frac{N_{\rm lim}}{T \cdot N_p \cdot \bar\sigma_{\rm IBD} \cdot \epsilon_{\rm sig}},
  \label{eq:upperbd}
  \end{equation}
where $N_{\rm lim}$ is the upper bound on the number of events for the given operation time $T$. The IBD cross section at $E_\nu=24.3$~MeV is used for the averaged cross section $\bar\sigma_{\rm IBD}$. In the following, we consider the sensitivity at HK. 
The signal efficiency assumed in this study ($\epsilon_{\rm sig}$) is taken from the former SK analysis~\citep[][]{2021PhRvD.104l2002A}, as was done in our previous study~\citep[][]{2022ApJ...937...30A}, which is around 20\%--30\% depending on energy. Incidentally, higher efficiency may be expected thanks to the better sensors used in HK.
As is done at SK, we assume that the IBD reaction of $\bar\nu_e$ is identified by the coincidence of the Cherenkov light emitted from the positron and the 2.2 MeV $\gamma$-ray emitted from neutron capture on hydrogen. This method is called neutron tagging and the signal efficiency becomes not very high due to the low energy of this $\gamma$-ray.

For a certain confidence level (C.L.), $N_{\rm lim}$ is obtained as the excess of the observation over the background expectation, where the statistical and systematic uncertainties of the background events are taken into account. The background at HK in higher energies ($E_\nu > 17.3$~MeV) mainly stems from atmospheric neutrinos and is classified into two categories: neutral-current quasielastic (NCQE) interactions and others (non-NCQE). In the NCQE interactions, a neutrino often knocks a neutron out of an oxygen nucleus, where the $\gamma$-ray emitted from the deexcitation of the residual nucleus and the knocked-out neutron mimic the positron signal from IBD. The size of the NCQE background assumed in the present study is scaled from that of the past SK experiment \citep[][]{2021PhRvD.104l2002A}. The systematic uncertainty of the NCQE background is assumed to decrease year by year \citep[see Table~1 of][]{2022ApJ...937...30A} due to expected efforts in accelerator neutrino and nuclear experiments \citep[][]{2019PhRvD.100k2009A,PhysRevC.109.014620,2024arXiv240515366T}. Since the systematic uncertainties would be improved by further efforts, our assumptions are conservative.

As for the non-NCQE background, we consider the following interactions: charged-current interaction of atmospheric $\bar\nu_e$, which produces a positron and a neutron as in IBD, charged-current muon neutrino interaction, and some of neutral-current interactions involving a low-energy pion. In the latter two interactions, a visible positron from the decay of an invisible muon becomes the background in the DSNB search because a neutron produced by the parent reaction may be tagged or, even without a neutron, an accidental coincidence with noise might occur. The size and systematic uncertainty of the non-NCQE backgrounds are again extrapolated from the SK experiment and conservative assumptions are adopted in this study. In addition, we also take into account the background due to the spallation of oxygen nuclei induced by energetic atmospheric muons, which produces radioactive isotopes and leads to misidentification as IBD events. HK is expected to suffer from the four times higher spallation background event rate per volume compared to SK due to the shallower depth of its construction site \citep[][]{2018arXiv180504163H}. Not only do the isotopes with $\beta$+$n$ decay, such as $^9$Li, increase but also the likelihood of accidental coincidences becomes higher. We then increase both $^9$Li and accidental backgrounds by a factor of four compared to the ones in \citet{2022ApJ...937...30A}\footnote{This is rather a conservative estimation because the accidental background in the energy range of the present analysis is not necessarily made by spallation but partially by atmospheric events as well.}. While this estimation may change with further detailed studies based on full simulations, the major spallation contribution in the current energy range is an accidental background which can be well understood from data-driven studies as performed at SK. Therefore, the presented sensitivity estimation could be regarded as conservative.

Using $\Phi_{\rm lim}$ obtained by eq.~(\ref{eq:upperbd}), we evaluate the operation time required to detect the DSNB at HK for our models. This is shown in Fig.~\ref{fig:sensitivity} as a function of $f_{\rm BHSN}$ for different cases and C.L. We found that the DSNB detection would be achieved within at most $\sim$6~yr at 2$\sigma$ and $\sim$20~yr at 3$\sigma$, and including the BH-forming SNe reduces the required operation time in any case. Furthermore, the impact of BH-forming SNe is significant in some cases; the operation time required for 3$\sigma$ is shortened by half with $f_{\rm BHSN}=0.2$ for the model of fallback induced BH formation, NH, and MD14 SFR. Otherwise, the upper bound on $f_{\rm BHSN}$ would be provided within the expected operational period of HK.

%% The "ht!" tells LaTeX to put the figure "here" first, at the "top" next
%% and to override the normal way of calculating a float position
%\begin{figure}[ht!]
%\plotone{samplefig.png}
%\caption{The cost for an author to publish an article has trended downward
%over time. This figure shows the average cost of an article from 1990 to 2020 in 2021 adjusted dollars. 
%\label{fig:general}}
%\end{figure}

\section{Summary and Discussion} \label{sec:discuss}
In this study, we have constructed a new DSNB model that includes the contribution of BH-forming SNe, which lead to both a successful SN explosion and BH formation simultaneously. According to studies on Galactic chemical evolution and nucleosynthesis, the population of BH-forming SNe is implied to be non-negligible in accounting for the observed abundance of some heavy elements. Since the detailed dynamics and neutrino emission of BH-forming SNe are uncertain, we have considered two extreme cases: fallback-induced BH formation and prompt BH formation. In the first scenario, a longer duration until BH formation ensures significant neutrino emission and the total energy of emitted neutrinos in our model is consistent with the binding energy of a maximum-mass NS. On the other hand, in the second scenario, the shorter duration results in reduced neutrino emission. The rates of successful SNe (the sum of ordinary core-collapse SNe and BH-forming SNe) and failed SNe (BH formation without SN explosions) have been based on the model of Galactic chemical evolution in \citet{2023MNRAS.518.3475T}. As a result, we have found that the contribution of BH-forming SNe enhances the flux and event rate of the DSNB at high energies ($E_\nu>17.3$~MeV). In particular, the impacts are the largest in the case of fallback-induced BH formation with neutrino oscillation in NH since the fallback mass accretion onto a PNS emits a much larger amount of $\bar\nu_e$ than $\nu_x$. In this case, the expected event rate at $E_\nu>17.3$~MeV doubles if the fraction of BH-forming SNe is $f_{\rm BHSN}=0.5$. Furthermore, with $f_{\rm BHSN} = 0.2$, the operation time required for $3\sigma$ detection at HK is shortened by half, assuming MD14 SFR. Similarly, in other cases, the required operation time is also reduced due to the contribution of BH-forming SNe.

Concerning the treatment of BH-forming SNe, there is room for further improvement. In this paper, we have avoided specifying the dynamics of BH-forming SNe and instead focused on discussing two extreme cases of neutrino emission. In actual conditions, the amount of neutrino emission may lie between these two extremes or vary widely. While numerical examples are still limited, the model of \citet{2023ApJ...957...68B} has a short time to BH formation of less than 2~s, which can be considered to be similar to our prompt BH formation model. Incidentally, \citet{2021ApJ...909..169K} show that even failed SNe may take around 10 s to form a BH. The time to BH formation is heavily dependent on the efficiency of fallback mass accretion, which is significantly determined by the structure of the progenitor. Therefore, it is worthwhile to investigate the neutrino emissions from BH-forming SNe using the same progenitor models employed in studies of nucleosynthesis and chemical evolution.

In the present study, we assume that BH-forming SNe reside within the mass range of 8--$18M_\odot$ according to the observed implication \citep{2009ARA&A..47...63S,2015PASA...32...16S}. In contrast, the arguments based on nucleosynthesis/chemical evolution have been done under the hypothesis that the progenitor masses of BH-forming SNe would be larger than $20M_\odot$ \citep[e.g.,][]{2006ApJ...653.1145K,2023MNRAS.524.6295P}. Thus, there is a clear inconsistency between the two.
While our results depend mainly on the fraction of BH-forming SNe, i.e., $f_{\rm BHSN}$, for the given rate of successful SNe, i.e., $R_{\rm SN}$, regardless of their progenitor masses, it is worthwhile to discuss how the observed upper mass bound ($18M_\odot$) can be reconciled with their masses ($>20M_\odot$) implied from the theoretical argument based on nucleosynthesis.

One possible explanation is the observational bias. Many BH-forming SNe may be unnoticed if they exhibit systematically lower peak luminosities than ordinary core-collapse SNe. This possibility is quite plausible for the case of faint SNe \citep[][]{2006NuPhA.777..424N}. Even for hypernovae, such a non-detection is possible owing to their jet-like explosions;~in most cases, the jets do not direct to us and the corresponding SNe seem to exhibit low brightness. Another possible solution is the metallicity-dependent frequency of BH-forming SNe;~they exclusively emerge in a low-metallicity environment, which is provided by the limited regions in the local Universe. This possibility is in particular expected for hypernovae because their progenitors are considered to be fast-rotating massive stars \citep[e.g.,][]{1998Natur.395..672I} and a low metallicity helps to retain enough angular momentum \citep[e.g.,][]{2006ApJ...637..914W}. In addition, hypernovae could be closely connected to long $\gamma$-ray bursts \citep{1998Natur.395..670G}, whose emergence is indeed biased toward low-metallicity ($Z\lesssim 0.3-0.5Z_\odot$) galaxies \citep{2006Natur.441..463F, 2015A&A...581A.102V}. These arguments propose that the redshift evolution of metallicity for individual galaxies could be one of the key factors including the IMF for counting the DSNB flux, as done by \citet{2015ApJ...804...75N}.

%% Also note that the akcnowlodgment environment does not support long amounts of text. If you have a lot of people and institutions to acknowledge, do not use this command. Instead, create a new \section{Acknowledgments}.
\begin{acknowledgments}
This work is supported by Grants-in-Aid for Scientific Research (JP18H01258, JP20K03973, JP23H00132, JP24K00632, JP24K07021), Grant-in-Aid for Scientific Research on Innovative Areas (JP19H05811), and Grant-in-Aid for Transformative Research Areas (JP24H02245) from the Ministry of Education, Culture, Sports, Science and Technology (MEXT), Japan, and by NSF Grant No. PHY-2309967. In this work, numerical computations were partially performed on the supercomputers at Research Center for Nuclear Physics (RCNP) in Osaka University.
\end{acknowledgments}

\bibliography{sample631}{}
\bibliographystyle{aasjournal}



\end{document}